\documentstyle[twoside,amssymb,fleqn,espcrc2]{article}

\newcommand{\AmS}{{\protect\the\textfont2
  A\kern-.1667em\lower.5ex\hbox{M}\kern-.125emS}}

\def\R{{\Bbb{R}}} 
\def\Z{{\Bbb{Z}}}
\def\be{\begin{equation}}\def\ee{\end{equation}}
\def\CH{{\cal H}}\def\CL{{\cal L}}\def\CP{{\cal P}} \def\CM{{\cal M}}
\def\0{\over } \def\6{\partial }
\def\({\left(} \def\){\right)} \def\<{\langle } \def\>{\rangle }
 
\let\lra=\leftrightarrow 
 
\def\wt{\widetilde}
\def\d{\delta}

\title{2d quantum dilaton gravity as/versus finite dimensional quantum 
mechanical systems}

\author{T. Strobl\address{Institut f\"ur Theoretische Physik E,
        RWTH--Aachen, \\ 
        Sommerfeldstr. 26-28, D-52056 Aachen, Germany}}

\begin{document}

\maketitle

In this talk we will deal with quantum aspects of generalized 2d
dilaton theories. The classical features of these models were
discussed already in the talk by Thomas Kl\"osch \cite{Thomas}. We
will also refer to \cite{Thomas} for the Lagrangian, Eq.\ 
(\cite{Thomas}.1), which governs the dynamics for our 2d metric $g$,
the scalar dilaton field $\Phi$, and, if $K \not\equiv 0$, the
Yang--Mills connection $A=A^{YM}$.  It contains four functions
$U,V,W,K$ to be fixed so as to define the system, allowing for a
simultaneous treatment of a whole family of models.

My talk will have three parts: First, I will summarize what one
might call the ``Chern--Simons''--formulation of generalized 2d
dilaton gravity \cite{TK1,Brief}. Second, I will deal with the
Hamiltonian quantization of these models \cite{Brief,TK4}. 
As one of the results of this analysis we will obtain the
spectrum of the mass operator, finding it to be 
sensitive to the signature of the theory as well as to the choice
of functions  $U,V,W,K$  in (\cite{Thomas}.1).
Third, I will make some remarks on the statistical mechanical
entropy that one obtains for the 2d models \cite{Kunst} 
when applying the approach of Carlip as well as 
Balachandran, Chandar and Momen \cite{Carlip}.

Although we deal with a two--dimensional field theory, on all of
the above three occasions one finds finite dimensional particle
systems to play a central role. Before going into further
details in separate sections below, I intend to outline the main
idea of how these relations come about:\newline 
{\bf 1)} Given a Lagrangian (\cite{Thomas}.1) with, 
for presentational reasons, $K \equiv 0$ ($\Rightarrow$ no Yang-Mills fields)
we will associate to it an {\em auxiliary}\/ ${\Bbb{R}}^3$  with
coordinates $X^i$, $i = 1,2,3$, that is equipped with a Poisson
bracket $\{X^i,X^j\}=\CP^{ij}(X)$. Here the ($X^i$--dependent)
matrix $\CP^{ij}$ is determined uniquely by the three
``potentials'' $U,V,W$.  For dimensional reasons $\R^3$ 
cannot be symplectic. However, the
$\R^3$ foliates (stratifies) into two--dimensional submanifolds
which are symplectic.  Correspondingly, these ``{\em symplectic
leaves}\/'' may be regarded as phase spaces of fictitious point
particles (although in general these phase spaces will have non--trivial
topology and not be a cotangent bundle). It turns out
that this $\R^3$ equipped with the Poisson bracket $\CP^{ij}$
can serve as a {\em target space}\/ of a 2d $\sigma$--model
which is {\em equivalent}\/ to the original theory described by
(\cite{Thomas}.1), while providing powerful techniques not available in the
metrical formulation.\newline  
{\bf 2)} 
Like any gravitational system the Lagrangian (\cite{Thomas}.1) is invariant
under diffeomorphisms.  As demonstrated in the previous talk
\cite{Thomas} (and references therein), for a given topology of
the spacetime manifold $\CM$ the space of solutions to the field
equations modulo diffeomorphisms is finite dimensional only.
This is characteristic for a topological field theory.  For $\CM
\sim S^1 \times \R$ and $K\equiv 0$, in particular, the solution space was
found to be two--dimensional, one of the two parameters being
the ``mass'' $M$ of the spacetime. In a Hamiltonian treatment of
the model (\cite{Thomas}.1) a symplectic reduction leads to a phase space
that is equivalent to this solution space.
Correspondingly we are again left with a two--dimensional
symplectic manifold (orbifold), which this time is the reduced
(i.e.\ physical) phase space RPS of the dilaton gravity
theory. This makes the 2d theory appear as a point particle
model. However, in general the RPS has a non--trivial topology.
Similarly, in a Dirac quantization the only continuous parameter
the  physical wave functions $\Psi_{ph}$ depend on is found to
be $M$ (in an appropriate polarization).  But again, in general
there are also further discrete labels $m,l$, $\Psi_{ph}= \Psi_{ph}(M,
m,l)$, which are {\em intertwined}\/ with $M$ in a 
nasty manner. While this shall be made more transparent in the
Sec.\ 2 below, let me mention here only that the
discrete labels are a  relic of the higher dimension where the
theory was defined before the reduction. In those cases where
they are present, they are the obstacle of regarding the 2d
system to be identical to a standard point particle system on a
line.\newline
{\bf 3)} Carlip and Balachandran et al.\ regard
spacetimes $\CM$ with boundary $\6\CM$. Identifying  $\6\CM$
with the (stretched) horizon of a black hole, its entropy  is
suggested to stem from tracing over quantized boundary modes. At
the example of 2+1 gravity in its Chern--Simons formulation
Carlip developed a recipe of how to induce from the bulk action
on  $\CM$ an action that governs the dynamics of the boundary
modes on $\6\CM$. Given our formulation of (\cite{Thomas}.1), we will be
able to adapt this recipe to the general class of 1+1 gravity
theories. Note that since now $\CM$ is 1+1 dimensional, its
boundary will be 0+1 dimensional, i.e.\ here the boundary
modes {\em are}\/ point particles. They are 
distinct from the bulk degrees of freedom certainly and thus
have to be distinguished from the finite dimensional modes
encountered in the two items above. Still, we will find a very 
nice relation to the fictitious particles on the target space; 
in fact, in a certain sense, 
the latter become alive and real by the approach of \cite{Carlip}. 

\section{``Chern--Simons'' formulation of generalized 2d dilaton gravity}
One of the major steps in the realm of 2+1 gravity was its
reformulation in terms of a Chern--Simons gauge theory, \be \int d^3x
\sqrt{|g|} \, \left(R+\Lambda\right) \lra \int tr \left(A \, dA + 
{2 \0 3} A^3\right) \,.  \ee Here the one--forms $A_i$, $i$ running over
the Lie algebra of the gauge group, have been identified with the
vielbein and the spin connection of an Einstein--Cartan formulation of
the gravity theory. A similarly powerful reformulation of the general
class of gravity models (\cite{Thomas}.1) exists in terms of Poisson
$\sigma$-models \cite{TK1,Brief}: \be L[g,\Phi,A^{YM}] \lra \int_\CM
A_i \wedge dX^i + {1 \0 2} \CP^{ij} A_i \wedge A_j \, . \label{equiv}
\ee We comment on this relation for the case that $K=W=0$, $U=\Phi$,
$V$ arbitrary, differentiable: Then $i,j =1,2,3$, again the $A$'s are
zweibein and spin connection, $A_i = \left(e^1, e^2,
\omega^1{}_2\right)$, while $X^3 \equiv \Phi$, and $X^1$ and $X^2$
have been introduced to ensure zero torsion in an Einstein--Cartan
formulation. The antisymmetric $3\times 3$--matrix $\CP$ is defined by
$\CP^{12}=V(X^3)$, $\CP^{13}=-X^2$, and $\CP^{23}=\pm X^1$, where the
two signs correspond to Euclidean and Lorentzian signature of the
gravity theory, respectively. Two decisive observations: 1) Regarding
$X^i$ as coordinates in an $\R^3$ (target space) and the lower index $i$ at
$A_i\equiv A_{i\mu}dx^\mu$ as a one--form index on this space, the
action (\ref{equiv}) is covariant with respect to ``coordinate
changes'' $X^i \to \widetilde X^i(X)$. 2) The matrix $\CP$ satisfies 
$ \CP^{il} \6\CP^{jk} /\partial X^l\,  + cycl.(i,j,k) = 0$ so that indeed
$\{ X^i, X^j \} := \CP^{ij}$ defines a Poisson bracket. 

The main technical advantage of the new formulation consists in
the existence of the following powerful tools: Locally the
Poisson bracket (in the target space) allows for
Casimir--Darboux coordinates. In the above example they may be
chosen as $\widetilde X^i := (M,\varphi,X^3)$ with
\be M = \pm (X^1)^2 +(X^2)^2 + 2 \int^{X^3}\! V(u)du \,\, , \label{M} \ee
while 
$\varphi = \arctan (X^2/X^1)$ and $\varphi = \ln(X^1+ X^2)$ for Euclidean and 
Lorentzian signature, respectively. Due to the simplification of the 
Poisson tensor in these coordinates the r.h.s.\ of  
(\ref{equiv}) {\em  trivializes locally}\/ to $\int A_{\wt i}
\wedge d \wt X^i + A_{\wt 2} \wedge A_{\wt 3}$ where $A_{\wt i}
\equiv A_j \6 X^j/\6 \wt X^i$. {\em Global}\/ information is restored 
by taking into account the topology of the symplectic leaves, which 
coincide with the connected components of the level surfaces of the 
Casimir function (\ref{M}).\footnote{At values of $M$ marking a transition 
in topology of the leaves, the connected components of the 
level surfaces $M=const.$ may still consist of several symplectic leaves.} 

These tools facilitate  even the analysis for a $\,$
$\CP^{ij}$ linear in $X$, in which case (\ref{equiv}) 
takes the form of an ordinary $BF$--gauge theory (cf., e.g.,
\cite{Brief,Jackiw}). 

\section{Hamiltonian quantization}
The r.h.s.\ of (\ref{equiv}) is already in first order form, so its
Hamiltonian structure is determined readily. Denoting the spacetime
coordinates with $r,t$, where $r$ may be a coordinate on either $S^1$
or $\R$, $X^i(r)$ becomes conjugate to $A_{jr}(\bar r)$, i.e.\ 
$\{X^i(r),A_{jr}(\bar r)\}=\d^i_j \d(r - \bar r)$, while the $A_{jt}(r)$
become Lagrange multipliers for the first class constraints $G^i(r)
\equiv {X^i}'+ \CP^{ij}(X) \, A_{jr} \approx 0$. Clearly the field
theoretic Poisson bracket has nothing to do with the one on the target
space and thus should be distinguished sharply from it.
\subsection{Reduced phase space}
Here we have to look at the quotient space$\( G^i(r) \approx
0\) / \{G^i(r), \cdot\}=:$ RPS. Taking into account boundary conditions, RPS
coincides topologically with the space of solutions to the field
equations modulo diffeomorphisms (if one excludes incomplete
spacetimes with kinks, to be precise). Thus in the case of periodic
boundary conditions in $r$ a direct comparison with the solution space
for $\CM = S^1 \times \R$, summarized in \cite{Thomas}, is
accessible. Indeed this solution space 
was found to be two--dimensional, one parameter
being a ``mass''--parameter $M$, which in fact coincides with the
(constant) value of (\ref{M}) on $\CM$, while the other one, which we
shall call $P$ here,  was given
a geometrical interpretation at the beginning of Sec.\ 3.1 and in
Sec.\ 3.3 of \cite{Thomas}. It may be shown that 
$P=\oint A_{\wt 1} dr$. Correspondingly the induced Poisson bracket 
on the RPS may be calculated, yielding $\{M,P\}=1$.
In the simply connected case  $\CM = \R^2$, on the other
hand, the solution space was found to be one--dimensional in
\cite{Thomas,TK1}, being parametrized by $M$. Here
$P=\int_{r_{min}}^{r_{max}} A_{\wt 1} dr$ becomes a 
second gauge--invariant parameter upon appropriate
choices of boundary conditions at  $r=r_{min,max}$; it then yields
the asymptotic Killing time difference of the $(t=0)$--hypersurface 
\cite{Kuchar}. So, in any case the RPS is two-dimensional. However,
generically its topology is quite
non--trivial. Depending on the choice of the potentials $U,V,W,K$, one
may run into qualitatively different situations:

{\em Simpler cases:}\/ This occurs for those potentials where the
integer $n$ defined in Sec.\ 1.2 of \cite{Thomas} is at most one, or,
equivalently, whenever {\em all}\/ the symplectic leaves in the target space
are simply connected.  Spherically symmetric 4d gravity as well as
string inspired gravity belong to this class of models. Here RPS
$=\R^2$ with globally conjugate variables $M \in \R$ and $P\in \R$.
Thus the RPS is equivalent to the one of a point particle on the line.
Correspondingly, up to unitary equivalence, $\Psi=\Psi(M)$, $P \to
-i\hbar d/dM$, and $\Psi \in \CH=\CL^2(\R,d M)$.

{\em Generic cases:}\/ Whenever $n_{max} \ge 2$, which occurs, 
 e.g., whenever $V$ is not positive or negative definite in the example
$W=K=0$, $U=\Phi$ above.  Here the topology of RPS is non--trivial,
since along the range of $M$ the value of the integer $n$ will change. A
typical scenario: For $M < M_1$ and $M>M_2$ one has $n=0$ with a
completely homogenous (or stationary) Penrose--diagram (cf.\ Fig.\ 1
in \cite{Thomas}); besides $M$ these solutions are labelled by one further
continuous parameter $P \in \R$.  For $M_1 < M < M_2$, however, $n=2$
and beside $P \in \R$ there is an additional {\em discrete}\/ label
$\l$ of the solutions, the integer patch number in the fundamental
region, cf.\ Fig.\ 5 of \cite{Thomas}. Thus now $\Psi=\Psi(M,l)$, but
$l$ may be non--zero {\em only}\/ within a certain range of $M$.
Correspondingly, we do not know the inner product for these wave
functions yet and quantization seems ambiguous in these cases.

The above analysis was valid for Lorentzian signature. For Euclidean
signature there are choices of the potentials such that $P$ takes
values from a finite interval only. In such cases the spectrum of the
conjugate variable $M$ will be discrete.

\subsection{Dirac quantization}
We have to solve the functional 
equations
\be 
 \left[{X^i}'(r) + i\hbar \CP^{ij}(X(r)) {\d \0 \d X^j(r)}\right]
\Psi[X(r)]=0 
\,. \label{qcons} \ee
Given the above factor ordering, there are no anomalies and thus
(\ref{qcons}) is integrable locally. In the case of our
three--dimensional target space with the Poisson tensor defined
by the single function $V$, the local solution to (\ref{qcons})
has the form: $\Psi[X(r)]=\delta[M'(r)] \, \exp\( (i/\hbar)
\oint \alpha \) \Psi_0(M)$. Here $M$ is the functional 
defined in (\ref{M}) and the 
first factor yields a restriction to maps $X(r)$ lying 
entirely in a symplectic leaf $M=const$. Furthermore, $d\alpha
=\Omega$ where $\Omega$ is the
symplectic form on the respective leaf,  induced by the target
space Poisson bracket. As a result of the analysis \cite{Brief} for a  {\em
global}\/ solution to the quantum constraints, we
merely have to determine the topology of the level surfaces
(\ref{M}). Let $r \sim r + 2\pi$ (periodic boundary conditions).
The topological invariants of relevance are $\pi_0(M=const)$,
$\pi_1(M=const)$, and $\pi_2(M=const)$. $\pi_0 \neq 0$ and/or
$\pi_1 \neq 0$ lead to an additional discrete label in $\Psi_0$:
$\Psi_0 \to \Psi_0(M,m,l)$, $m \in  \pi_0$, $l \in \pi_1$. While
in the Lorentzian case $m$ may often be disposed of by
the discrete Lorentz symmetry $X^1 \lra -X^1$, the integer
$l$ is precisely the quantity encountered in \cite{Thomas} and 
the previous subsection  as patch number.  Here it is the winding number of
the argument loop of $\Psi$ around non--simply connected pieces
of the leaf $M=const$. Since the topology of the leaves will in
general change with the value of $M$, we again get this
entanglement of $l$ with $M$. In our present context 
$\pi_2(M=const) \neq 0$ iff (\ref{M}) is compact. For an $M$ with a
compact surface (\ref{M}), there exists a physical wave
function only if 
\be \oint \Omega = 2\pi n \hbar  \quad, \quad n \in \Z \, \, ,\label{O} 
\ee where the integral is taken over all of the leaf. 
Note that in general this may yield a discrete spectrum for $M$.
Be warned also not to confuse $\Omega$ with the symplectic form
of the field theory; it is the one on the leaf $M=const.$ in the
target space. The condition (\ref{O}) arises as an
additional {\em global}\/ integrability condition to the quantum
constraints (\ref{qcons}). As a general fact, for the Lorentzian signature 
$\pi_2(\mbox{leaves})=0$, while for the Euclidean signature 
$\pi_1(\mbox{leaves})=0$. Irrespective of the signature, in the 
{\em simpler cases}\/ (defined as in Sec.\ 2.1) 
$\pi_1=\pi_2=0$ and we again obtain $\Psi_{phys}
\lra \Psi_0(M)$.  A first example with  non--trivial $\pi$'s is provided by
$V(\Phi)=\Phi$ (2d deSitter gravity): In the Lorentzian 
theory the leaves (\ref{M}) are not simply connected for 
$M>0$,  while they are for $M<0$.  In the Euclidean theory
$\pi_2 = \Z$ and (\ref{O}) leads to $M = n^2 \hbar^2/4$;
$|n\rangle$ gives a basis for the physical wave functions in
this case. Generically, however, the spectrum of $M$ is mostly
continuous also in the Euclidean case 
\cite{Brief,TK4}.

\section{Statistical mechanical entropy}
Locally a classical solution of our gravity system (with $K\equiv 0$) 
is determined by the
value of $M$, which is a constant over spacetime. Now, given a
classical solution $\CM$ with boundary $\6 \CM$ and mass $M$, Carlip's
recipe may be applied to obtain the boundary action and from it the
phase space of the boundary particles. As a result of the analysis
\cite{Kunst} one
finds that this edge particle phase space is {\em isomorphic}\/ to the
symplectic leaf singled out from the Poisson bracket in the target
space by the value of $M$ and Eq.\ (\ref{M}). Applying the method of
geometric quantization to this phase space of the edge modes, we
obtain Eq.\ (\ref{O}) as a consistency condition.  Note that the
bulk modes are left unquantized in this approach; still, by quantizing
the analogue of Carlip's boundary modes, one obtains precisely the same
spectrum of $M$ as in a Dirac quantization of the bulk modes. The
attempt to determine  the entropy of the respective black hole by a
simple counting of physical ``boundary states'' may be successful only
in the case of a compact phase space. Only in that case the Hilbert
space of the boundary modes is finite dimensional, and, according to 
(\ref{O}), its dimension may be approximated very well by the integer $n$. 
The result, $S \sim \ln n$, gives, however, only about the logarithm of
what one expects for the entropy $S$ from other, semiclassical considerations. 
For most choices of $U,V,W$, however, the symplectic leaves are
non--compact in the Lorentzian {\em and}\/ the Euclidean theory, and 
a simple counting like this does not make sense.

\end{document}